\documentclass[aps,prb,preprint,superscriptaddress,showpacs,floatfix]{revtex4}

\usepackage{graphicx,color}

\graphicspath{{figs/}}

\begin{document}
\title{Intrinsic Negative Poisson's Ratio for Single-Layer Graphene}
\author{Jin-Wu Jiang}
    \altaffiliation{Corresponding author: jiangjinwu@shu.edu.cn}
    \affiliation{Shanghai Institute of Applied Mathematics and Mechanics, Shanghai Key Laboratory of Mechanics in Energy Engineering, Shanghai University, Shanghai 200072, People's Republic of China}

\author{Tienchong Chang}
    \affiliation{Shanghai Institute of Applied Mathematics and Mechanics, Shanghai Key Laboratory of Mechanics in Energy Engineering, Shanghai University, Shanghai 200072, People's Republic of China}

\author{Xingming Guo}
    \affiliation{Shanghai Institute of Applied Mathematics and Mechanics, Shanghai Key Laboratory of Mechanics in Energy Engineering, Shanghai University, Shanghai 200072, People's Republic of China}

\author{Harold S. Park}
    \affiliation{Department of Mechanical Engineering, Boston University, Boston, Massachusetts 02215, USA}

\date{\today}
\begin{abstract}

Negative Poisson's ratio (NPR) materials have drawn significant interest because the enhanced toughness, shear resistance and vibration absorption that typically are seen in auxetic materials may enable a range of novel applications.  In this work, we report that single-layer graphene exhibits an intrinsic NPR, which is robust and independent of its size and temperature.  The NPR arises due to the interplay between two intrinsic deformation pathways (one with positive Poisson's ratio, the other with NPR), which correspond to the bond stretching and angle bending interactions in graphene. We propose an energy-based deformation pathway criteria, which predicts that the pathway with NPR has lower energy and thus becomes the dominant deformation mode when graphene is stretched by a strain above 6\%, resulting in the NPR phenomenon.

\end{abstract}

\pacs{61.48.Gh, 62.20.-x}
\keywords{Graphene, Negative Poisson's Ratio}

\maketitle
\pagebreak


When materials are stretched in a particular direction, they typically contract in the directions orthogonal to the applied loading.  The Poisson's ratio ($\nu$) is the material property that characterizes this behavior, and it is typically positive, in the range of $0.2< \nu < 0.5$ for most engineering materials.  A negative Poisson's ratio (NPR), in which the material expands in the directions orthogonal to the applied loading, is allowed by classical elasticity theory, which sets a range of $-1< \nu < 0.5$ for the Poisson's ratio in an isotropic three-dimensional material.\cite{LandauLD}

Because most materials have a positive Poisson's ratio, significant efforts have been made to discover NPR materials starting with the seminal work of Lakes in 1987.\cite{LakesR1987sci} In this experiment, the NPR was induced by the permanent compression of a conventional low-density open cell polymer foam, which was explained by the re-entrant configuration of the cell. Since then, many works have achieved the NPR through structural engineering or via composite structures.\cite{RothenburgL1991nat,LakesR1993adm,BaughmanRH1993nat,EvansKE2000adm,YangW2004jmsci,RaviralaN2007jms,LethbridgeZAD2010am,BertoldiK2010am,GreavesGN2011nm,AldersonK2012pssb,clausenAM2015}  Materials with NPR have become known as auxetic, as coined by Evans.\cite{EvansKE1991Endeavour}

Besides the NPR phenomena found in specifically engineered structures, the NPR has also been found to be intrinsic for some materials.  For example, some cubic elemental (both FCC and BCC) metals have intrinsic NPR along a non-axial, i.e. $\langle110\rangle$ direction.\cite{MilsteinF1979prb,BaughmanRH1998nat} The Poisson's ratio for FCC metals can be negative along some principal directions by controlling the transverse loading.\cite{hoPSSB2016b}  NPR was found to be intrinsic to single-layer black phosphorus due to its puckered configuration, which leads to NPR in the out-of-plane direction.\cite{JiangJW2014bpnpr} NPR was also predicted to be intrinsic for few-layer orthorhombic arsenic using first-principles calculations.\cite{HanJ2015ape}

While most existing works are on bulk auxetic structures, some theoretical works have recently emerged predicting NPR in nanomaterials, through a range of different mechanisms. For example, Yao et al. investigated the possibility of inducing NPR in carbon nanotubes, though significant, and likely non-physical changes to either the structural parameters or bonding strengths were found to be necessary for the NPR to appear.\cite{YaoYT2008pssb} The NPR for metal nanoplates was found due to a surface-induced phase transformation.\cite{HoDT2014nc} Furthermore, the NPR for graphene has been found through various means, including due to the rippling curvature induced by entropic effects at very high (1700 K) temperatures\cite{ZakharchenkoKV}, by creating periodic porous graphene structures\cite{hoPSSB2016a}, and introducing many vacancy defects\cite{GrimaJN2015adm}.  More recently, two of the current authors found that the compressive edge stress-induced warping of the free edges can cause NPR in graphene nanoribbons with widths smaller than 10~nm.\cite{JiangJW2016npr_fbc}

The salient point in the above literature review is that it is important to uncover intrinsic NPR in specific materials, including graphene, the most widely studied material since its discovery in 2004. However, the literature to-date has shown that the NPR phenomenon can only be observed in graphene after specific engineering of its structure, such as thermally-induced ripples, vacancy defects, or free edges. More specifically, it is still unclear whether the NPR is an intrinsic property for graphene.

In this letter, we reveal that NPR is intrinsic to single-layer graphene, and is independent of its size and temperature.  More specifically, the Poisson's ratio evolves from positive to negative when the applied tensile strain exceeds about 6\%. We find that this NPR is due to the interplay between two fundamental deformation pathways, which we term PW-I and PW-II, and which correspond to two characteristic interactions in graphene, i.e., the bond stretching and angle bending. The PW-I deformation mode yields a positive Poisson's ratio, while the PW-II deformation mode results in a NPR. We therefore propose a pathway-based energy criterion, which predicts that the PW-II mode becomes more important than the PW-I mode and dominates the deformation mechanism of graphene for strains larger than 6\%. Consequently, the Poisson's ratio becomes negative when the applied tensile strain is larger than 6\%.

\begin{figure}[htpb]
  \begin{center}
    \scalebox{1.4}[1.4]{\includegraphics[width=8cm]{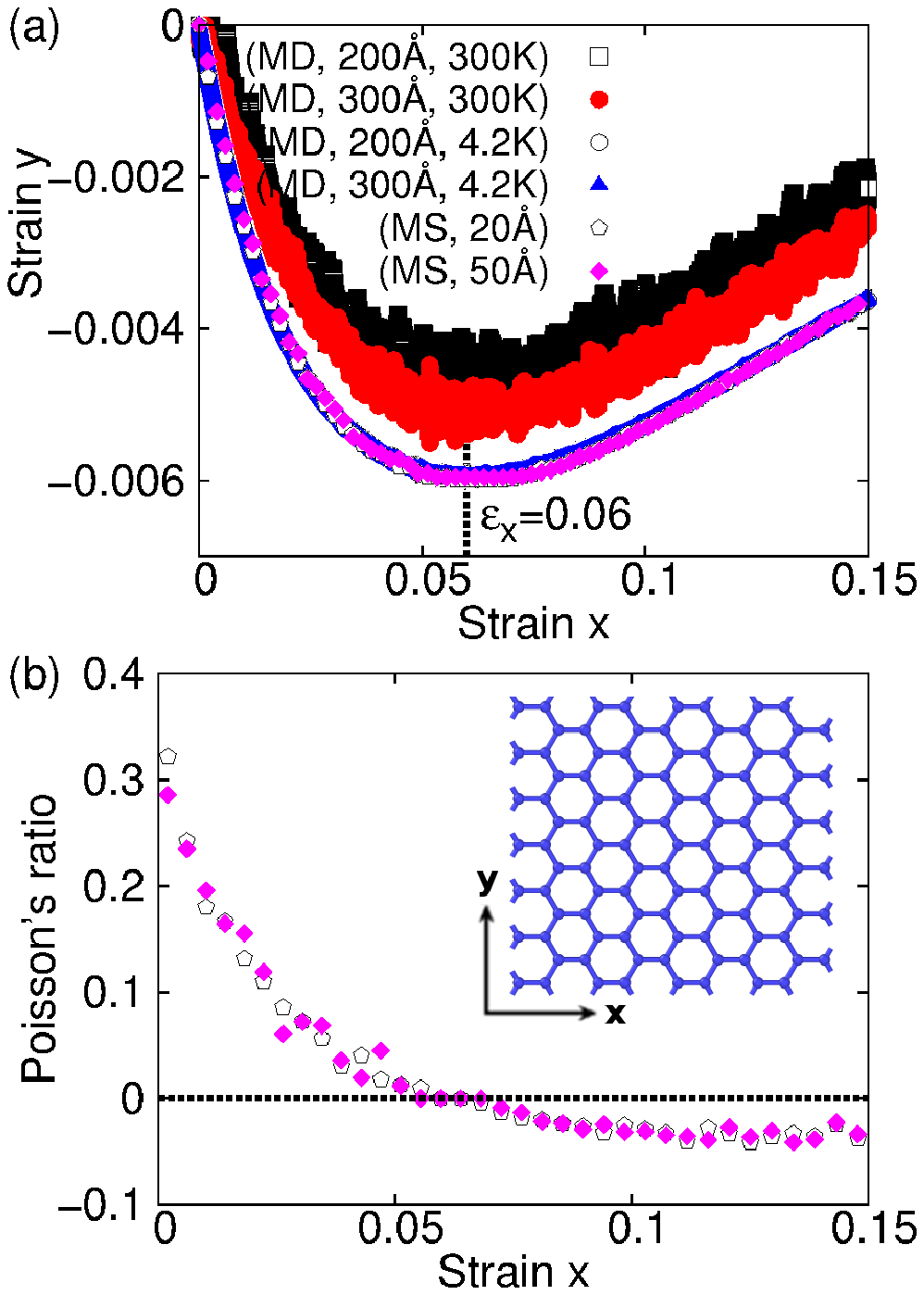}}
  \end{center}
  \caption{(Color online) Negative Poisson's ratio in graphene. (a) The resultant strain $\epsilon_y$ versus the applied strain $\epsilon_x$. A robust valley point exists at $\epsilon_x=0.06$ in all curves for varying simulation parameters using both MD or MS approaches. (b) The Poisson's ratio extracted from the MS results in (a) through $\nu=-\partial \epsilon_y/\partial \epsilon_x$, which is negative for $\epsilon_x>0.06$.}
  \label{fig_poisson}
\end{figure}

\textbf{Results.} The crystal structure for single-layer graphene is shown in the inset of Figure~\ref{fig_poisson}(b). Periodic boundary conditions are applied in both x and y-directions such that our studies, and the properties we report, represent those of bulk graphene without edge effects. The carbon-carbon interactions are described by the Brenner potential,\cite{brennerJPCM2002} which has been widely used to study the mechanical response of graphene.\cite{MoY2009nat} The structure is stretched in the x (armchair)-direction while graphene is allowed to be fully relaxed in the y (zigzag)-direction, using both molecular dynamics (MD) or molecular statics (MS) simulations. For the MD simulations, the standard Newton equations of motion are integrated in time using the velocity Verlet algorithm with a time step of 1~{fs}, which is small enough to maintain energy conservation during the MD simulations. This time step is also small enough to accurately discretize the atomic trajectory corresponding to the highest-frequency vibration modes in graphene (with frequency around\cite{JiangJW2008prb} $4.8\times 10^{13}$~Hz). For the MS simulations, the conjugate gradient algorithm is used for energy minimization. Simulations are performed using the publicly available simulation code LAMMPS~\cite{PlimptonSJ}, while the OVITO package is used for visualization~\cite{ovito}.


Figure~\ref{fig_poisson}~(a) shows the resultant strain $\epsilon_y$ in the y-direction in graphene that is stretched by $\epsilon_x$ in the x-direction. The x-axis is along the horizontal direction, while the y-axis is in the vertical direction as shown in the inset of Figure~\ref{fig_poisson}~(b). The resultant strain in the y-direction is computed by $\epsilon_{y} = \frac{L_{y}-L_{y0}}{L_{y0}}$ with $L_{y0}$ and $L_y$ as the initial and deformed lengths in the y-direction. We simulate the tensile deformation of graphene using both MD and MS simulations. MD simulations are carried out for square graphene of size $L=200$~{\AA} and 300~{\AA} at 4.2~K and 300~K. MS simulations are performed for graphene of dimension $L=20$~{\AA} and 50~{\AA}.  

As shown in Figure~\ref{fig_poisson}~(a), there is a robust valley point around $\epsilon_x=0.06$ (6\%) in all of these curves. This valley point indicates that the Poisson's ratio, calculated by\cite{HoDT2014nc} $\nu=-\partial \epsilon_y/\partial \epsilon_x$, is positive for $\epsilon_x<0.06$ but becomes negative for $\epsilon_x>0.06$. These results demonstrate that the NPR is robust, as it is observed for both low and room temperature conditions, as well as for all structural sizes we have considered. We note that graphene is highly stretchable and has been stretched in a wide strain range experimentally. A strain up to 0.15 has been applied on graphene to measure the nonlinear stiffness\cite{LeeC2008sci} or manipulate its rippled structure.\cite{BaoW2009nn} Furthermore, recent experiments reported a uniaxial strain up to 0.1, which can be controlled in a reversible and nondestructive manner in graphene.\cite{GarzaHHP2014nl}  Thus, the critical strain of $\epsilon_x=0.06$ we find has been achieved in contemporary experiments on graphene, so that theoretical results in the present work are experimentally verifiable. The Poisson's ratio is about 0.3 at $\epsilon_x\approx 0$, which agrees with a recent numerical result with the realistic interatomic potential LCBOPII.\cite{LosJH2006prl}

Figure~\ref{fig_poisson}~(a) shows that results from MD simulations at 4.2~K coincide with the results from MS simulations, so we will concentrate on the MS simulation results for the rest of this paper. Figure~\ref{fig_poisson}~(b) shows the strain dependence for the Poisson's ratio extracted from these two curves from MS simulations in panel (a). It explicitly shows that the Poisson's ratio is negative for $\epsilon_x>0.06$.

We note that there have been previous reports of NPR in graphene.  Specifically, large numbers of vacancy defects\cite{GrimaJN2015adm} or patterning periodic porous structures\cite{hoPSSB2016a} for bulk graphene or compressive edge stress-induced warping in graphene ribbons are three different mechanisms to achieve the NPR in graphene.\cite{JiangJW2016npr_fbc} The Poisson's ratio for graphene can also be driven into the negative regime by thermally induced ripples at high temperatures above 1700~K.\cite{ZakharchenkoKV} In contrast, the NPR revealed in the present work represents an intrinsic material property for single-layer graphene.

\begin{figure}[htpb]
  \begin{center}
    \scalebox{1.8}[1.8]{\includegraphics[width=8cm]{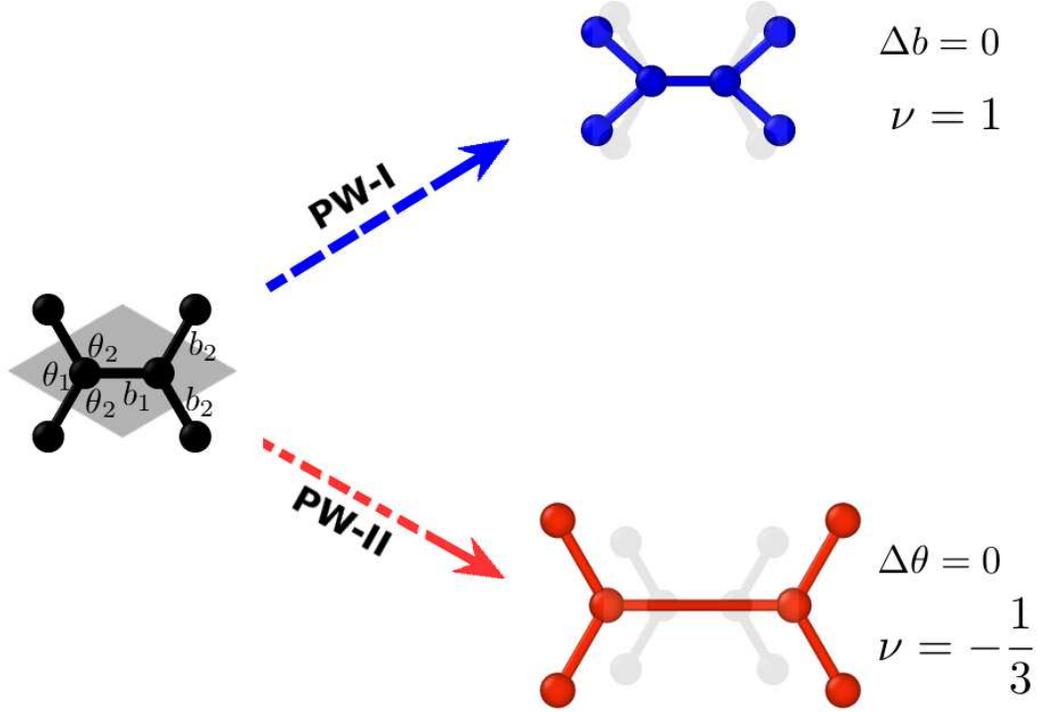}}
  \end{center}
  \caption{(Color online) Two typical ideal deformation pathways during the tensile deformation of graphene. The left atom cluster (black online) is stretched along the horizontal x-direction. The parallelogram gray area indicates the unit cell. PW-I (blue online): carbon-carbon bond lengths remain constant ($\Delta b=0$), while angles are altered to accommodate the external strain, which results in a Poisson's ratio of $\nu=1$. PW-II (red online): angles are unchanged and bond lengths are elongated to accommodate the external tension, resulting in a NPR of $\nu=-1/3$. The lighter shades show the undeformed structure.}
  \label{fig_pathway}
\end{figure}

\textbf{Discussion.} To explore the underlying mechanism for the intrinsic NPR, we first illustrate two major deformation modes for the tensile deformation of graphene in Figure~\ref{fig_pathway}. These two deformation modes are fundamental deformation modes corresponding to the bond stretching and angle bending interactions;\cite{ChangT2003jmps} i.e., $V_{b} = \frac{K_{b}}{2}\left(b-b_{0}\right)^{2}$ and $V_{\theta} = \frac{K_{\theta}}{2}\left(\theta-\theta_{0}\right)^{2}$, where $b$ is the bond length and $\theta$ is the angle in the deformed graphene, and $\theta_0=120^{\circ}$ and $b_0=1.42$~{\AA} are material constants related to undeformed graphene. $K_{b}$ is the force constant that characterizes the resistance to bond stretching, and so a larger value of $K_{b}$ indicates a stiffer bond. $K_{\theta}$ characterizes the resistance to angle bending, and so a larger value of $K_{\theta}$ means a larger resistance to angle bending deformations. The values of these force constants can be obtained by using the value of the Young's modulus and the Poisson's ratio of graphene.\cite{ChangT2003jmps} The bond stretching and angle bending are two major interaction terms in graphene, especially in graphene without out-of-plane deformation. We note that the bond stretching and angle bending interactions were used to derived analytic expressions for the Poisson's ratio in carbon nanotubes.\cite{ChangT2003jmps,ShenL2004prb,ChangTC2005apl,YaoYT2008pssb} The analytic expressions illustrate the dependence of the Poisson's ratio on geometrical parameters and force constants. For example, Yao et al. performed a speculative investigation on the possibility of NPR for carbon nanotubes by varying one parameter (or ratio of parameters) while holding all other parameters unchanged.\cite{YaoYT2008pssb}  However, it was determined that significant, and likely non-physical changes to the geometry or material constants would be necessary for the NPR to appear in CNTs.

The overall deformation process for graphene depends on the competition between the bond stretching and angle bending interactions. For $V_b\gg V_\theta$, the bonds are too stiff to be stretched ($\Delta b\approx 0$), so only the bond angles will change during the tension of graphene. This type of deformation will be referred to as the PW-I deformation mode, as shown in the top (blue online) of Figure~\ref{fig_pathway}. For $V_b\ll V_\theta$, the bond angles cannot be changed ($\Delta \theta\approx 0$), so the bond lengths will be stretched to accommodate the applied tension. This type of deformation will be referred to as the PW-II deformation mode, as shown in the bottom (red online) of Figure~\ref{fig_pathway}.

The Poisson's ratio corresponding to the PW-I and PW-II deformation modes can be derived by simple algebra. The unit cell is displayed by the parallelogram gray area in the left configuration in Figure~\ref{fig_pathway}. The size of the cell in Figure~\ref{fig_pathway} in the x and y-directions are
\begin{eqnarray}
L_{x} & = & 2\left(b_{1}+b_{2}\cos\frac{\theta_1}{2}\right)\\
L_{y} & = & 2b_{2}\sin\frac{\theta_{1}}{2},
\end{eqnarray}
which yields
\begin{eqnarray}
dL_{x} & = & 2\left(db_{1}+db_{2}\cos\frac{\theta_1}{2}-\frac{b_{2}}{2}\sin\frac{\theta_{1}}{2}d\theta_{1}\right)\\
dL_{y} & = & 2\left(db_{2}\sin\frac{\theta_{1}}{2}+\frac{b_{2}}{2}\cos\frac{\theta_{1}}{2}d\theta_{1}\right).
\end{eqnarray}
As a result, the Poisson's ratio is
\begin{eqnarray}
\nu & = & -\frac{\epsilon_{y}}{\epsilon_{x}}=-\frac{dL_{y}/L_{y}}{dL_{x}/L_{x}}\nonumber\\
 & = & -\frac{b_{1}+b_{2}\cos\frac{\theta_{1}}{2}}{b_{2}\sin\frac{\theta_{1}}{2}}\times\frac{db_{2}\sin\frac{\theta_{1}}{2}+\frac{b_{2}}{2}\cos\frac{\theta_{1}}{2}d\theta_{1}}{db_{1}+db_{2}\cos\frac{\theta_{1}}{2}-\frac{b_{2}}{2}\sin\frac{\theta_{1}}{2}d\theta_{1}},
\label{eq_nu}
\end{eqnarray}
where $b_{1}=b_{2}=b_{0}$ and $\theta_{1}=\theta_{2}=\theta_{0}$ for undeformed graphene. We note that, for small strains, the definition of the Poisson's ratio in equation~(\ref{eq_nu}) is consistent with the numerical formula used in the above to extract the Poisson's ratio in Figure~\ref{fig_poisson}~(b), because $\epsilon_y$ and $\epsilon_x$ have a linear relationship for small strains.

For the PW-I mode, we have $\Delta b_{1}\approx0$, $\Delta b_{2}\approx0$, and $\Delta\theta_{1}\not=0$. As a result, we obtain the Poisson's ratio for PW-I mode as $\nu=1$ from equation~(\ref{eq_nu}). For the PW-II mode, we have $\Delta\theta_{1}\approx0$, and the force equilibrium condition leads to\cite{ChangT2003jmps} $db_{2}=\frac{1}{2}db_{1}\cos\frac{\theta_{1}}{2}$. Hence, the Poisson's ratio for the PW-II mode is $\nu=-1/3$ from equation~(\ref{eq_nu}). It is interesting to note that the Poisson's ratio ($\nu=-1/3$) for the PW-II mode coincides with the expectations of the self consistent screening approximation.\cite{LosJH2006prl}

\begin{figure}[htpb]
  \begin{center}
    \scalebox{1.5}[1.5]{\includegraphics[width=8cm]{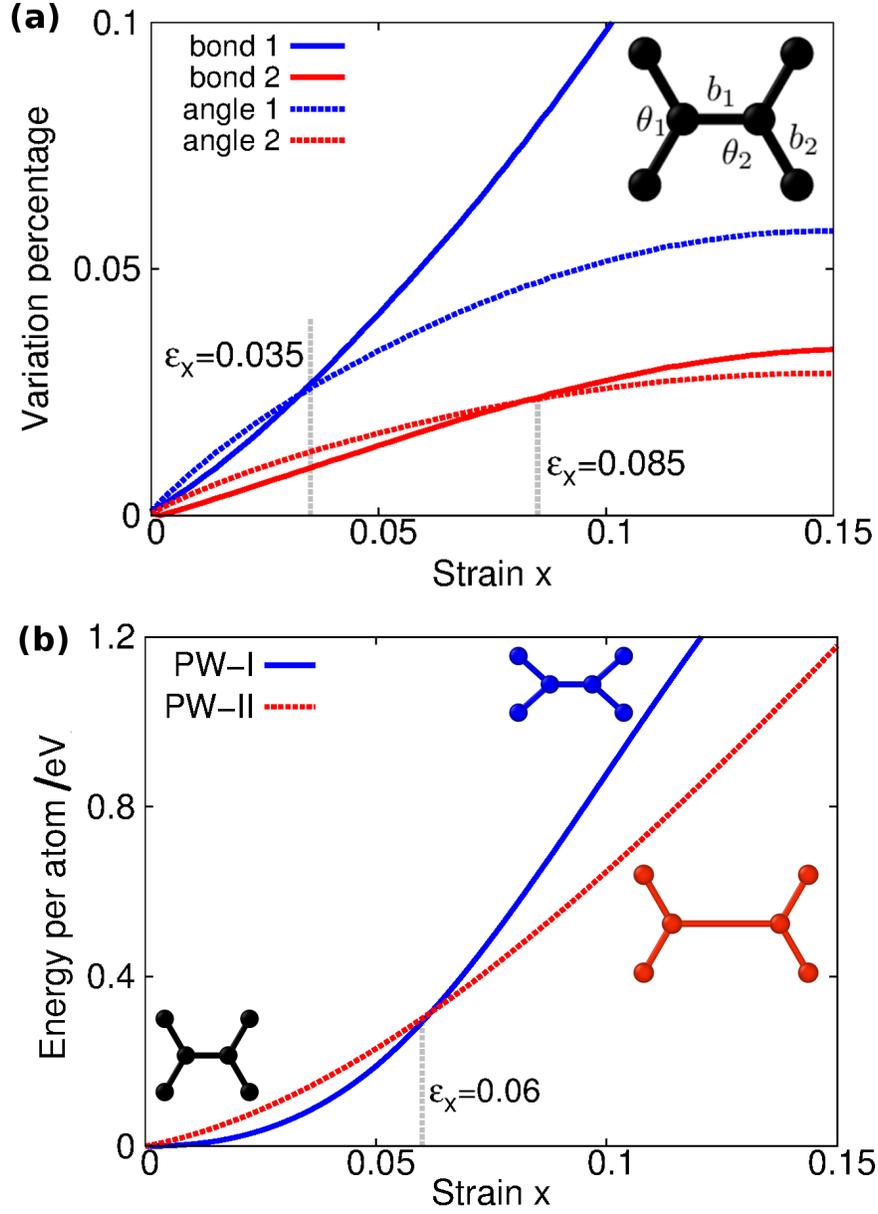}}
  \end{center}
  \caption{(Color online) Pathway energy criteria for PW-I and PW-II deformation modes. (a) The variation of key geometrical parameters (angles $\theta$ and bond lengths $b$) in graphene. The y-axis shows the relative variation, i.e., $\Delta b/b_0$ or $\Delta \theta/\theta_0$. (b) Pathway energy curve for PW-I and PW-II deformation modes.  The curves show a crossover at $\epsilon_x=0.06$, which predicts a transition from PW-I mode (positive Poisson's ratio) to PW-II mode (negative Poisson's ratio) during the tensile deformation of graphene. Left bottom inset (black online) shows the undeformed structure. Top inset (blue online) displays the PW-I deformed structure. Right inset (red online) is the PW-II deformed structure.}
  \label{fig_bondangle_energy}
\end{figure}

Figure~\ref{fig_bondangle_energy}~(a) illustrates the interplay between the PW-I and PW-II deformation modes during the tensile deformation of graphene.  Specifically, it shows the absolute value of the relative variation of the angles $\theta_{1}$ and $\theta_{2}$, and bond lengths $b_{1}$ and $b_{2}$; we note that the change in angle $\theta_{1}$ is negative in stretched graphene. For $\epsilon_x<0.035$, the variations of angles $\theta_{1}$ and $\theta_{2}$ are larger than the variations of bonds $b_{1}$ and $b_{2}$, respectively, which indicates PW-I to be the dominant deformation mode for graphene subject to small uniaxial tensile strains. For $0.035<\epsilon_x<0.085$, the variation of angle $\theta_{1}$ becomes less than the variation of bond $b_{1}$, while the variation of angle $\theta_{2}$ is still larger than the variation of bond $b_{2}$, which implies a competition between the PW-I and PW-II deformation modes for moderate strains. For $\epsilon_x>0.085$, variations for both bonds are larger than the variations of angles, so PW-II overcomes PW-I to be the dominant deformation mode for large tensile strains. Hence, the value of the Poisson's ratio will decrease with increasing strain, and will become negative at some critical strain between $[0.035, 0.085]$ when PW-II dominates the deformation process. The critical strain of 0.06 for the NPR in Figure~\ref{fig_poisson} falls in this strain range.

We perform an energy-based analysis, shown in Figure~\ref{fig_bondangle_energy}~(b), of the PW-I and PW-II deformation modes to gain further insight into the interplay between these two deformation modes governing the transition from positive to negative Poisson's ratio at $\epsilon_{x}=0.06$.  The energy curve is computed as follows. For PW-I, the structure is manually deformed corresponding to the PW-I mode shown in the top of Figure~\ref{fig_pathway}. We then calculate the potential energy of this deformed structure, which is higher than the potential energy of undeformed graphene. The energy curve shown in Figure~\ref{fig_bondangle_energy}~(b) is the difference between the potential energy per atom of the deformed and undeformed graphene structures. The energy curve for PW-II is computed similarly, where angular distortions are allowed while the bond lengths are kept constant.  Figure~\ref{fig_bondangle_energy}~(b) clearly shows a crossover around $\epsilon_x=0.06$ between the energy curves for PW-I and PW-II modes.

\textit{We thus propose a pathway energy based criteria: the tensile deformation process for graphene is governed by the deformation mode with lower pathway energy.} According to this criteria, PW-I mode will be the major deformation mode for $\epsilon_x<0.06$, in which the pathway energy for PW-I mode is lower than the pathway energy for PW-II mode. Similarly, the pathway energy criteria predicts the PW-II to be the major deformation mode for $\epsilon_x>0.06$, in which PW-II has lower pathway energy. We showed in Figure~\ref{fig_pathway} that the PW-I mode has a positive Poisson's ratio, while the PW-II mode has a NPR. As a consequence, the Poisson's ratio is positive for $\epsilon_x<0.06$, and will turn negative for $\epsilon_x>0.06$. This prediction is in excellent agreement with the numerical results in Figure~\ref{fig_poisson}, where the Poisson's ratio changes from positive to negative at $\epsilon_x=0.06$. We note that, to our knowledge, it is the first time the pathway energy criteria is proposed, which is based on the energetic competition for the two major in-plane deformation pathways. This criteria may be useful for future investigations into the mechanical properties of nano-materials similar as graphene.

\textbf{Conclusion.}  In conclusion, we performed both MD and MS simulations to investigate the Poisson's ratio of graphene in the strain range 0 $<$ $\epsilon$ $<$ 0.15.  We observed an intrinsic NPR for tensile strains exceeding $\epsilon_{x}=0.06$, which is independent of graphene's size and temperature.  The appearance of the NPR is a direct result of the interplay between the PW-I (with positive Poisson's ratio) and PW-II (with NPR) modes during the tensile deformation of graphene. These two deformation modes correspond to the fundamental bond stretching and angle bending interactions, which are the two major in-plane interaction components in graphene.  These results were further validated through a pathway energy criteria to predict positive or negative Poisson's ratio in graphene.  Using this simple model, we found that the pathway energy for the PW-II deformation mode becomes lower than the PW-I mode for graphene tensile by strain above 0.06, leading to NPR above this strain range.

\textbf{Acknowledgements} The work is supported by the Recruitment Program of Global Youth Experts of China, the National Natural Science Foundation of China (NSFC) under Grant Nos. 11504225, 11472163, 11425209, and the start-up funding from Shanghai University. HSP acknowledges the support of the Mechanical Engineering department at Boston University.

\textbf{Author contributions} J.W.J performed the calculations and discussed the results with T.C, X.G, and H.S.P. J.W.J and H.S.P co-wrote the paper. All authors comment on the paper.

\textbf{Competing financial interests} The authors declare no competing financial interests.


\providecommand{\latin}[1]{#1}
\providecommand*\mcitethebibliography{\thebibliography}
\csname @ifundefined\endcsname{endmcitethebibliography}
  {\let\endmcitethebibliography\endthebibliography}{}

\end{document}